# Experimental realization of coexisting states of rolled-up and wrinkled nanomembranes by strain and etching control

*Peter Cendula[1,*,†], Angelo Malachias[2,3], Christoph Deneke[1,2], Suwit Kiravittaya[1,§] and Oliver G. Schmidt[1]*

[1]Institute for Integrative Nanosciences, IFW Dresden, Helmholtzstrasse 20, D-01069 Dresden, Germany

[2] Laboratório Nacional de Nanotecnologia (LNNano), Rua Giuseppe Máximo Scolfaro 10000, 13083-100, Campinas, SP, Brazil

[3]Departamento de Física, Universidade Federal de Minas Gerais, CP 702, 30123-970 Belo Horizonte, MG, Brazil

KEYWORDS : rolled-up, wrinkles, crumples, x-ray diffraction, finite element method.

ABSTRACT

   Self-positioned nanomembranes such as rolled-up tubes and wrinkled thin films have been potential systems for a variety of applications and basic studies on elastic properties of nanometer-thick systems. Although there is a clear driving force towards elastic energy minimization in each system, the exploration of intermediate states where specific characteristics could be chosen by a slight modification of a processing parameter had not been experimentally realized. In this work, arrays of freestanding III-V nanomembranes (NM) supported on one edge and presenting a coexistence of these two main behaviors were obtained by design of strain conditions in the NMs and controlled selective etching of patterned substrates. As the etching process continues a mixture of wrinkled and rolled-up states is achieved. For very long etching times an onset of plastic cracks was observed in the points with localized stress. The well-defined morphological periodicity of the relaxed NMs was compared with finite element simulations of their elastic relaxation. The evolution of strain in the NMs with etching time was directly evaluated by X-ray diffraction, providing a comprehensive scenario of transitions among competing and coexisting strain states.

Self-positioning of strained thin films is a promising key technology for future applications. Although the current state of layer engineering strongly lies on the evolution of an initial flat state towards either a rolled-up or a wrinkled configuration, one can realize systems in which specially designed initial conditions allow for achieving either one single final state or a coexistence of both states, providing the necessary flexibility desired for the use of a single layered sample layout on any application type. In particular, the interest in wrinkling phenomena was recently revived through their applications in metrology,[1] nanochannels for sensing and manipulation[2] and stretchable electronics.[3] Different physical configurations lead to wrinkling patterns and the most frequently used is stiffing thin films on thick elastic foundation.[4,5] Another wrinkling configuration consists of subjecting films to boundary confinement,[6] which is observed in different size scales ranging from meter-long drapes to atomically-thin suspended graphene,[7] passing through systems of technological interest such as strained free-hanging thin films.[8–11] On the other hand, strained films under specific boundary constraints can also roll-up[12] into nano- and microtubes when sufficient strain gradient is present in the films.[13–16] Rolled-up nanotubes are engineered to microjets,[17,18] nanosprings,[19] ring resonators,[20–22] magnetic memory devices[23] and self-positioned micro-batteries.[24] Graphene layers were also shown to be suitable for actuator applications upon rolling.[25,26] We previously used elasticity theory to show that a competition between roll-up and wrinkling morphologies takes place at particular strain states,[11] but the experimental realization of competing roll-up and wrinkling relaxation is still lacking. Finally, even if only fundamental studies of material properties are concerned, methods involving nanomembranes transfer or overgrowth[27,28] can be considerably influenced by the competition of rolling-up and wrinkling for a given system.

In this work, we designed semiconductor strained nanomembranes (NM) and successfully obtained structures with coexisting rolled-up and wrinkled morphologies after their elastic relaxation. Semiconductors are ideally suited for such study since they allow precise strain engineering through control of alloy composition (here: Indium concentration) in the individual layers.[29] Strain gradient and average strain in the NMs were chosen to enable both rolling-up and wrinkling of the NMs. In such conditions, distinct strain states of the NM are locally obtained by controlling the etching time. We studied wavelength, strain and elastic energy of the observed morphologies by comparison with analytical expressions and numerical calculation with finite-element method (FEM). Average strain states and relaxation behavior were directly probed by synchrotron X-ray diffraction (XRD). The elastic energy results retrieved from these measurements were compared with scanning electron microscopy (SEM) observations and FEM strain analysis.

In order to induce a rolling-up tendency on a given NM bilayer the strain gradient $\Delta\varepsilon = |\varepsilon_2 - \varepsilon_1|$ between stacked epitaxial layers has to be equal or larger than 0.5%.[11,30] Therefore, we have used a $In_{0.1}Ga_{0.9}As$/GaAs bilayer of equal thickness $d_1= d_2=d$ with initial strains $\varepsilon_1 = -0.71\%$ and $\varepsilon_2 = 0\%$ (see Figure 1). The layers were grown by molecular beam epitaxy (MBE), on top of an etchant sensitive AlAs layer, deposited on a GaAs (001) substrate (see Refs. [30–32] for details). Etching depth $h$ of the AlAs layer in the samples was achieved by controlling time intervals of the immersion into a diluted HF solution. Directly after the etching, the samples were transferred to supercritical dryer to overcome deformation collapse due to capillary forces during drying in air,[33] exhibiting then the partially relaxed film morphologies explored in this work.

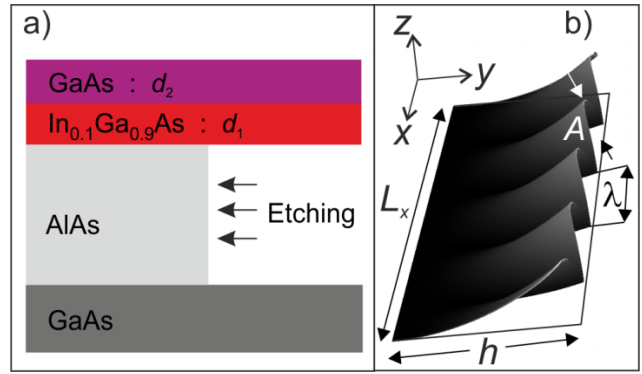

**Figure 1**. (a) Schematic representation of layer stack and etching of the AlAs sacrificial layer. (b) Simulation of a relaxed NM supported on the left edge along with initial flat shape (solid

rectangle) and coordinate system used. The variables discussed in the text are represented in the Figure 1.

The total elastic energy $E_0$ of our NM portion with length $L_x$ before any relaxation (induced by the etching of the sacrificial layer) is

$$E_0 = \frac{Y d_1 \varepsilon_1^2}{1-\nu} h L_x, \qquad (1)$$

Throughout this article, we assume a constant value of Young modulus $Y = 80$ GPa and Poisson ratio $\nu = 0.31$ for both layers in the NM. Following our previous results, the energy of the `pure rolled-up` NM (tube) with the final equilibrium radius $E_{R0}$ is given by [11]

$$\frac{E_{R0}}{E_0} = \frac{9-7\nu}{16}, \qquad (2)$$

and the equilibrium radius $R_0$ assuming a plane-strain condition (no longitudinal relaxation) is[14,30]

$$R_0 = \frac{4d}{-3\varepsilon_1(1+\nu)}. \qquad (3)$$

To allow wrinkling of the NM on the supported edge, sufficiently large average strain $\bar{\varepsilon} = (\varepsilon_1 + \varepsilon_2)/2$ should exist. Wrinkling onsets when a critical value of the governing parameter $\bar{\varepsilon} h^2/d^2$ is reached.[11,34] The periodicity (hereafter referred as wavelength) of wrinkles $\lambda$ without rolling-up (termed `pure wrinkles` in this article) should scale with $\lambda \cong 3.3h$ at the onset of buckling[9] and assume values of $\lambda \sim h^{0.65}$ [10] or $\lambda \cong 1.1h$ [34] in the post-buckling regime. Previous studies of wrinkles supported on one edge used mainly single strained layer.[8–10] We use a strained bilayer model in this article and therefore wrinkling mixed with rolling-up can develop different scaling behavior with respect to wavelength. Hence, considering the scenario described above we employ finite-element calculations to find wavelengths of relaxed structures from the elasticity theory.

We previously calculated a resulting morphology diagram showing rolled-up and wrinkled regimes as a function of strain gradient, average strain in the NM and etching depth.[11] Based on this diagram and the layers used in this work with $\bar{\varepsilon} = -0.35\%$, the preferential shape should be coexisting rolling-up and wrinkled independently of the etching depth. We will verify this prediction experimentally in this work.

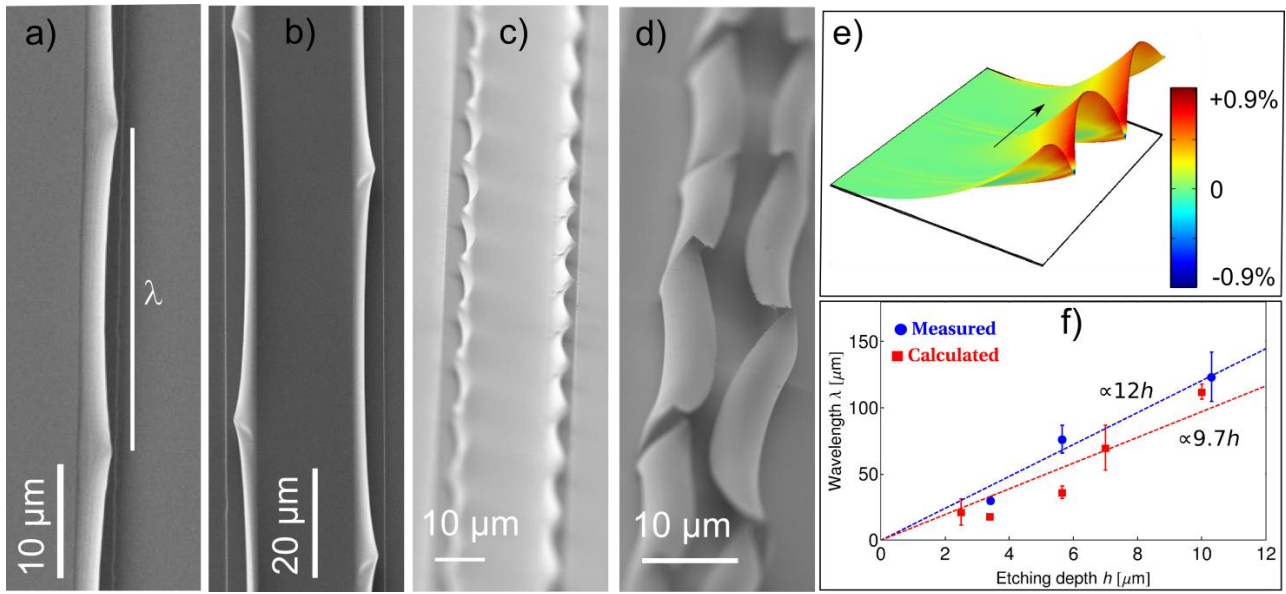

**Figure 2.** Morphology evolution for increasing etching time for sample batch A. SEM images are shown for: a) $h$ = 3.4 µm (top view); b) $h$ = 5.7 µm (top view); c) $h$ = 5.7 µm (side view), d) $h$ = 10.3 µm (side view). e) FEM simulation of longitudinal strain (direction of the arrow) for $h$ = 5.7 µm. f) Wavelength extracted from SEM images (see also Supporting Information Fig. S2-S8) and calculated from FEM simulations as a function of etching depth, along with fitted linear functions.

For sample batch A, thicknesses of the bilayer were chosen to be 20nm +20 nm and the etching rate in HF(1:900) solution was determined as 2.4 µm/min from SEM images (see supplementary material). The expected equilibrium tube radius is $R_0$ = 2.9 µm and total circumference of one tube rotation is $2\pi R_0$ = 18 µm, which is the required etching depth for usual rolling-up into tube (without wrinkling effects). Four SEM images depicting snapshots of intermixing rolled-up and wrinkled shapes as a function of etching depths are shown on Figures 2a-2d. For an etching depth $h$ = 3.4 µm (Fig. 2a), we observe a stage of bent NM corresponding rough to expected 1/6 of tube radius along with wrinkles of wavelength $\lambda$ = 30±3 µm on its edge. For etching depth $h$ = 5.7 µm, the sample is rolled-up to about 1/3 of the total rotation of the expected equilibrium tube. However, rolling-up is not ideal and wrinkles with wavelength $\lambda$ = 76±10 µm are also observed on tube surface, Fig. 2b and 2c. The sample with $h$ = 10.3 µm has rolled-up to about ½ of tube rotation with large wrinkles of wavelength $\lambda$ = 124±19 µm, as shown in Fig. 2d. The measured wavelength of wrinkles $\lambda$ as function of etching depth $h$ is summarized in Fig. 2f along with the wavelength extracted from FEM calculations of elastic relaxation of NMs.[35] The linear fits of measured and calculated wrinkle wavelength agree fairly with $\lambda \approx 12h$ and $\lambda \approx 9.7h$, respectively. A FEM calculated shape of the relaxed nanomembrane showing the strain in the longitudinal direction ($x$ axis) for $h$ = 5.7 µm is seen in Fig. 2e. The longitudinal strain values are close to zero near the supporting edge (etching front), but vary considerably along the wrinkled (free) edge of NM. In the larger concave wrinkle region, longitudinal strain is high tensile with magnitude of 0.9% and in the smaller convex wrinkle region longitudinal strain is high compressive also with magnitude of about -0.9%. The area of the convex region is very small compared to the area of the concave region and longitudinal strain is highly concentrated near punctual regions, which resemble a singularity similar to crumples in paper and thin sheets.[36–38]

In the second sample batch B, we used thinner membranes, with 10 nm (InGaAs) + 10 nm (GaAs) bilayer thickness. An etching rate 4.5 µm/min in HF(1:450) solution was determined from SEM images. The expected equilibrium tube radius is $R_0$ = 1.45 µm and circumference of one tube rotation is $2\pi R_0$ = 9 µm. SEM images of the evolving NM shape with increasing etching depth $h$ = 0.7 µm, 1.9µm, 2.7µm, 3.5µm and 7.1 µm are shown on Figs. 3a, 3b, 3c, 3d and 3e. If the NMs would roll into a tube, the predicted states of rotation would be fractions of 0.08, 0.20, 0.30, 0.40 and 0.80 of one tube rotation.

In Figure 3f, we used FEM to calculate the evolving NM shape. The longitudinal strain distribution (in original coordinate system) of sample with $h$ = 0.7 µm shows similar strain distribution in comparison to our results depicted in Fig. 2e. The longitudinal strain is close to zero near the supporting edge (etching front), but varies considerably along the wrinkled (free) edge of NM. In the larger concave wrinkle region, longitudinal strain is highly tensile with magnitude of 1.0%, while in the smaller convex wrinkle region longitudinal strain it is highly compressive also with magnitude of about -1.0%. The coexistence of the 'pure wrinkling' process with simultaneous rolling-up is also well visible in Fig. 3f in deflection from the initial plane (solid borders). The measured and calculated wavelength of wrinkles $\lambda$ as function of etching depth $h$ is summarized in Fig. 3g. The linear fits of measured and calculated wrinkle wavelength agree fairly with $\lambda \approx 9.3h$ and $\lambda \approx 6.5h$, respectively.

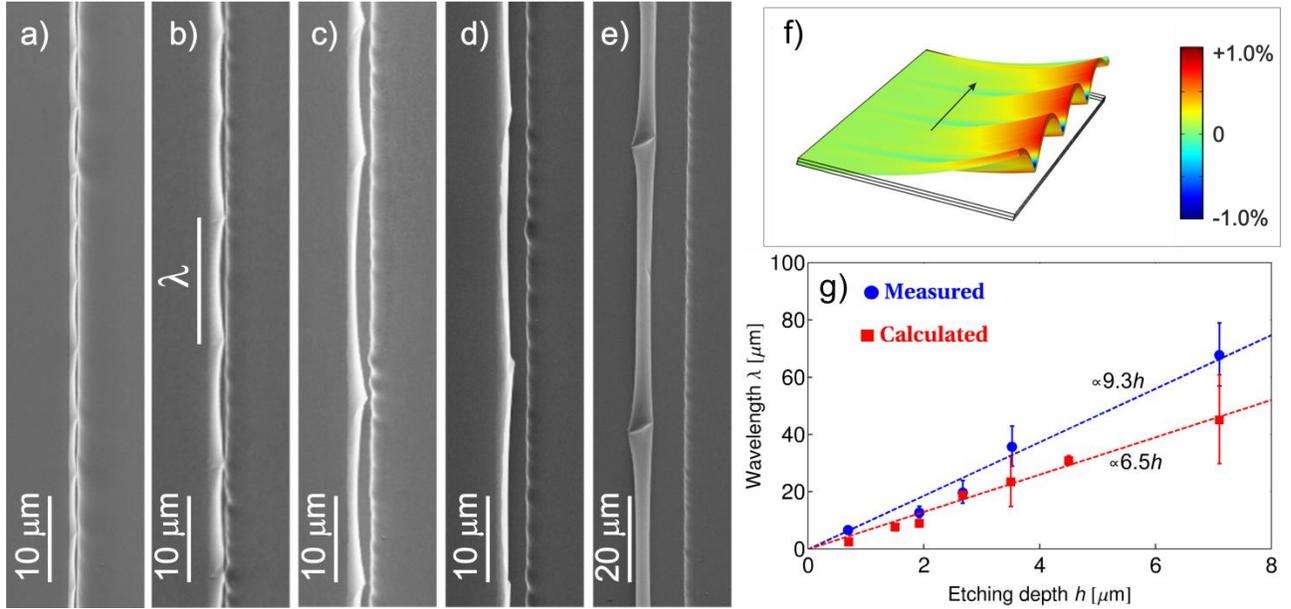

**Figure 3.** Morphology study of sample batch B. a)-e) Top view SEM images of various etching length $h = 0.7\mu m$, $1.9\mu m$, $2.7\mu m$, $3.5\mu m$ and $7.1 \mu m$ show wrinkles with increasing wavelength $\lambda$ (almost full tube for the large wrinkle in e)). f) FEM simulation of longitudinal strain (direction of the arrow) for $h = 0.7 \mu m$. g) Measured and calculated (FEM) wavelength of wrinkles $\lambda$ from a)-e) as a function of etching depth $h$, along with fitted linear functions.

We compare parameters depending on the elastic relaxation for the two sample batches A and B as a function of etching depth $h$. As first parameter, we plot the measured and calculated wavelength in Fig. 4a. The values from both batches show a roughly linear behavior with $h$ with $\lambda \cong 11h$ for measured wavelength and $\lambda \cong 8.7h$ for calculated wavelength from FEM. Both behaviors deviate substantially from scaling at wrinkling onset $\lambda \cong 3.3h$ [9,34] and scaling $\lambda \cong 1.1h$ in the post-buckling regime.[11,34] Therefore, additional effects, such as stress focusing, must play a considerable role during relaxation of strained bilayers, which were not taken into account in the analytical calculations. This is a consequence of the fact that rolling-up (bending) enforced by strain gradient of the NM is energetically favorable with respect to wrinkling (stretching) enforced by average strain in the NM.

The second parameter is the wrinkling amplitude plotted against etching depth in Fig. 4b. The wrinkle amplitude $A$ was extracted only from FEM calculations and it cannot be quantitatively extracted from SEM images. Both batches A and B also exhibit a roughly linear increase of amplitude $A$ as function of $h$, which is consistent with the linear scaling $\lambda \sim h$ obtained in the literature.[10,11,34]

Evaluating the total strain energy $E_{tot}$ of the relaxed NM portion from FEM calculations is useful in order to probe the quantitative degree of elastic relaxation of NMs. In Fig. 4c, we show the $E_{tot}$ as a function of the etching depth $h$. Despite of the existence of local minima for small $h$, $E_{tot}$ increases with increasing etching depth $h$ for both batches A and B. The increase in $E_{tot}$ with increasing $h$ indicates a less efficient elastic relaxation process that is caused by coexisting wrinkling and rolling behavior. It is useful to compare $E_{tot}$ with energy of rolled-up tube $E_{R0} \cong 0.43 E_0$ (eq. 2) and with the asymptotic energy of the wrinkled film $E_w \cong 0.6 E_0$.[11] For batch A and $h$ smaller than 5 μm, $E_{tot}$ is smaller than $E_{R0}$, implying the elastic relaxation of intermixed rolled-up and wrinkled morphology achieves a lower energetic state than only rolling-up. For $h$ larger than 5 μm, $E_{tot}$ is bigger than $E_{R0}$ and NM is forced to attain intermixed rolling-up and wrinkled morphology (due to both strain gradient and average strain present in the bilayer) even though it is an energetically less favorable state than only rolling-up. The larger values of wrinkling energy $E_w$ compared to the total energy calculated from FEM $E_{tot}$ of batches A and B confirms that elastic energy is relaxed more efficiently in FEM calculations than in the analytic calculation of wrinkling energy[11].

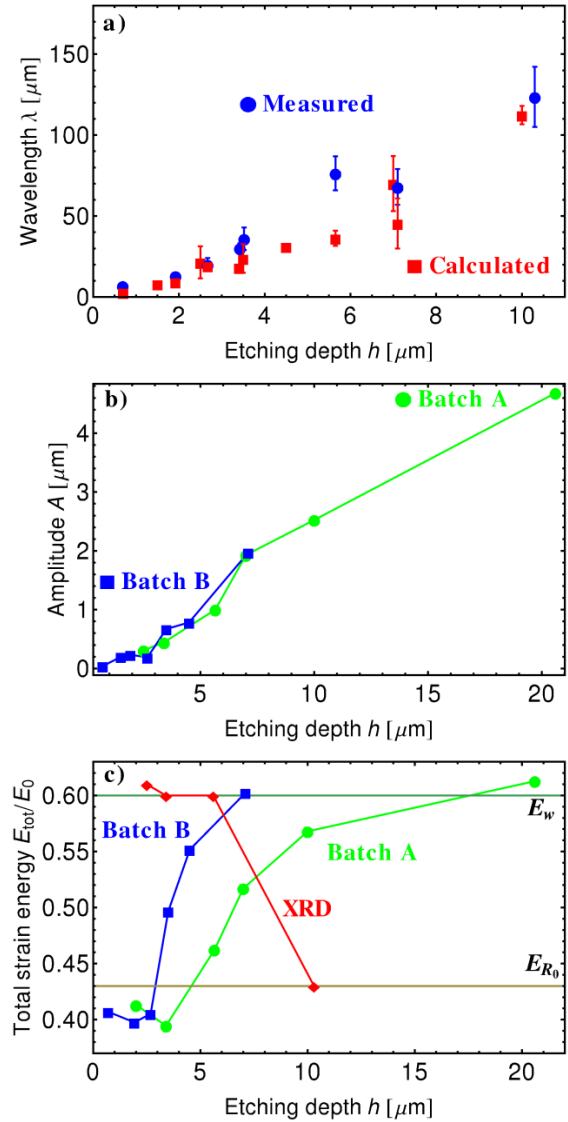

**Figure 4** a) Wavelength $\lambda$, b) calculated amplitude $A$ and c) total strain energy of relaxed NMs for samples from both batches A and B as a function of etching depth evaluated from FEM calculations (batch A and B) and XRD analysis for batch A.

Samples were etched for longer times than those shown on Fig. 2 or Fig. 3. In these cases cracked tubes with random periodicity were observed (see also Supporting Information Fig. S8), probably as a result of focused stress as observed for our samples with moderate etching time.

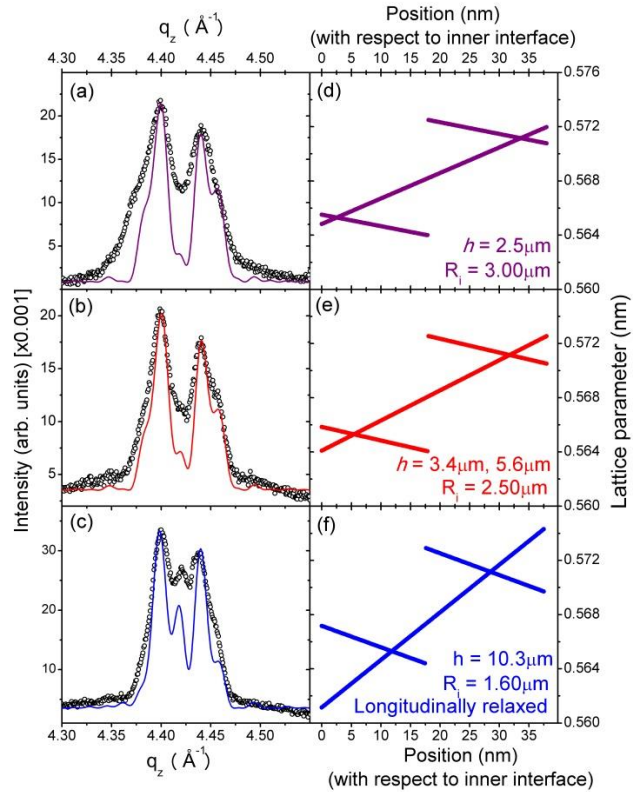

**Figure 5** XRD measurements and analysis for samples with $h = 2.5$ μm (top panels), $h = 3.4$ μm (middle panels) and $h = 10.3$ μm (lower panels). Left panels a), b), c) show diffraction measurements (dots) and fits with a kinematical model (solid lines)[39]. Right panels d), e), f) show the radial and tangential lattice parameter distributions along the layer thickness position retrieved from the XRD analysis.

Besides the simulations and microscopy results discussed above, strain status of rolled-up or wrinkled thin films are directly evaluated by X-ray diffraction. As depicted in detail in Ref. [39] a coplanar scattering geometry can be used to measure the radial lattice parameter of tube-shaped structures, providing a result that reflects the distribution of lattice parameter in the radial direction. Since the tubes behave as two-dimensional crystals, their scattering can be evaluated by mapping the vicinity of the GaAs (004) reflection with a de-tuning of the substrate angle of the order of 10° (here we used a de-tuning towards lower angles). By fitting the observed diffraction profiles with a kinematical model one can evaluate the local strain (with or without longitudinal relaxation), curvature and overall mosaic [39–41].

Figures 5a-5c shows the diffraction data collected at the LNLS (dots) for samples from the batch A and fits using kinematical theory (solid lines). On the Figures 5d-5f the tangential lattice parameter distribution (continuous line through the whole plot) and radial lattice parameter distribution (profiles with discontinuity in the middle) obtained from the simulations are shown. The diffraction data, fits and lattice profiles shown correspond to samples with etching length $h =$ 2.5 μm (Figs. 5a and 5d), $h = 3.4$ μm (Figs. 5b and 5e) and $h = 10.3$ μm (Figs. 5c and 5f). The dataset obtained for $h = 5.7$ μm is very similar to the data for $h = 3.4$ μm, and the profile used for fitting it is exactly the same. In order to fit all the curves, independently of the etching length the following parameters where fixed: i) the GaAs layer thickness was found to be 18.0(3) nm after rolling; ii) the InGaAs layer thickness was found to be 20.0(4) nm and; iii) the In content in the InGaAs layer was retrieved as 11%, very close to the nominal value.

The diffraction curves in the Figures 5a, 5b and 5c were plotted with the same $q$-scales ($x$-axis). By looking initially solely to the data points one clearly observes that the diffraction data extend over a larger $q$-range for the shorter etching lengths ($h = 2.5$ μm), and keep shrinking as etching length increases. Since the diffraction experiment only probes the scattering of a small angular section of the order of 0.5° of the tubes[39] the retrieved local radius must be regarded as the center of the radius distribution of the ensemble of objects (in all cases here incomplete tubes). Such radius distribution for the ensemble of objects is much larger for smaller etching length and narrows as the etching proceed, meaning that the distribution of radii along the tube axis (due to stress concentration points) is reducing. This is in agreement with the larger wavelength of wrinkles along the tube and consequently the reduction of the total amount (area density) of stress concentration points, as observed from microscopy and retrieved from FEM calculations.

The fits shown in the left panels of Fig. 5 allow the extraction of the best single-rolling parameters for each etching length. The equilibrium radius was found to vary near the expected radius for completely rolled tubes (about 2.50 µm). In order to fit the $h = 2.5$ µm curve, we had to increase the radius as a fitting parameter, which is seen as a larger separation between the peaks seen at the upper panel. This indicates that the layers are bending under a built-in strain which is not large enough to cause complete (and homogeneous) roll. Therefore, the hinged partial tubes rolled for $h = 2.5$ µm exhibit a larger average radius of 3.00(2) µm. The overall radius distribution cannot be easily depicted since there is an asymmetric distribution of the local radius, which reaches its minimum/maximum values near the stress concentration points. For the tubes from etched length $h = 3.4$ µm and $h = 5.6$ µm the average radius obtained of 2.50(2) µm is the one expected by energy minimization, and a consequence of the larger wavelength. The fit also has a better quality, indicating that most of the probed tube regions are within the expected curvature. In both cases – $h = 2.5$ µm and $h = 3.4$ µm – the longitudinal lattice parameter was found to remain at the GaAs bulk value, indicating that for the tube section probed local fluctuations of the longitudinal strain (Fig. 2e) lead to an average lattice parameter near the original flat layer in-plane registry.

For $h = 10.3$ µm on Fig. 5c and 5f the local curvature of the partially rolled structure varies longitudinally as well as tangentially (according to the distance from the etching front). The result is that for the sector probed by x-rays the radius becomes smaller since it is located at the tube side wall [39]. The retrieved radius is found to be of 1.60(3) μm, while in average the SEM images point out to 3.0(5) μm average radius. For this etching length, the SEM picture of Fig. 2d shows that the tubes start to crack. The plastic relaxation indicates that elastic energy can no longer be held only by wrinkling and plastic relaxation sets in. The reduced separation between cracks affects the longitudinal constraint with the substrate and induces a modification (relaxation) of the longitudinal lattice parameter, which is in agreement with the observed cracking of the tubes. As a consequence, not only the radius changes for this configuration but also the longitudinal lattice parameter reaches the average value between $a_{GaAs}$ and $a_{InGaAs}$ (for 10% In content).

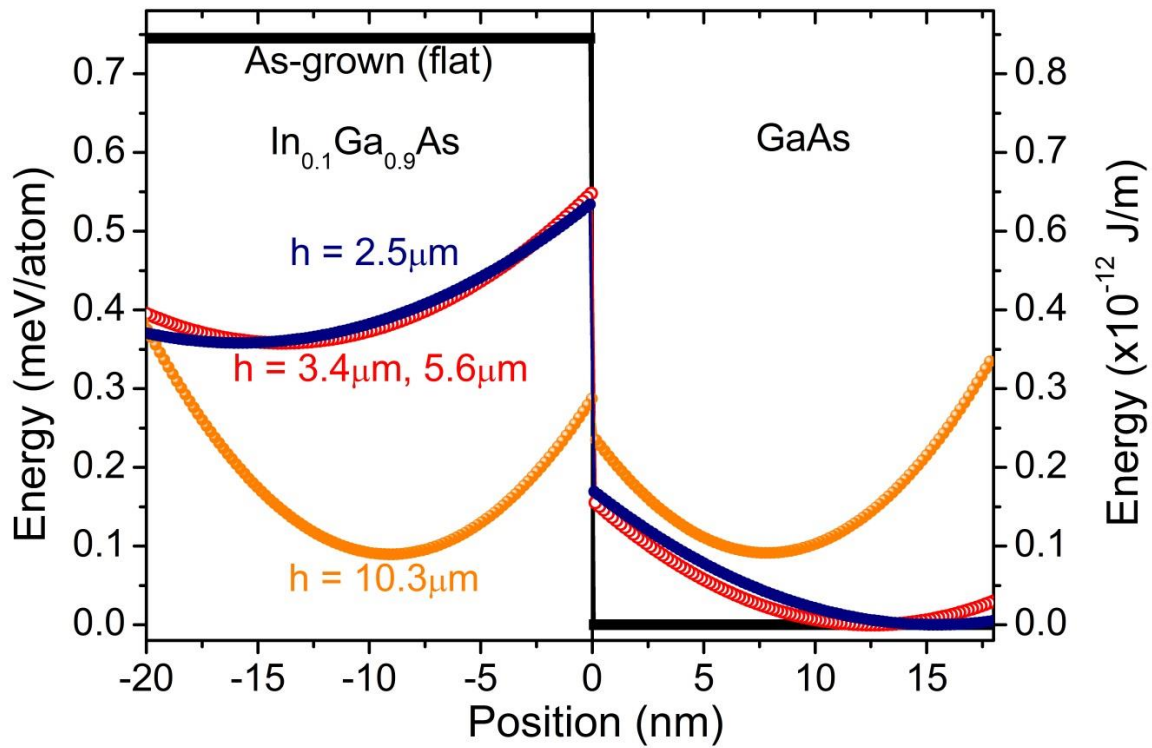

**Figure 6** Elastic energy per atom along the nanomembrane radial direction (position zero refers to the InGaAs/GaAs interface), extracted from XRD measurements for batch A for etching depths $h$ = 2.5 µm, 3.4 µm (5.6 µm) and 10.3 µm. A Joule/meter energy conversion is directly shown at the right y-axis.

The elastic energy per atom for samples from batch A was calculated from the difference of retrieved XRD lattice parameter profiles of Fig. 5 and bulk lattice parameters.[39] Figure 6 shows the initial strain status of the flat layers, where all elastic energy is stored in the InGaAs pseudomorphically strained layer, as well as the evolution of elastic energy for the etching depths analyzed here. For etching depths $h$ = 2.5 µm, 3.4 µm, 5.6 µm, wrinkling dominates the overall layer morphology and most of the elastic energy is still stored in the In-rich layer. After the onset of plastic relaxation (and consequently an equalization of longitudinal lattice parameter) a more homogeneous partition of the elastic energy takes place, as observed for etching depth $h$ = 10.3 µm. Integrating the XRD elastic energy per atom over thickness of the nanomembrane, we get values of total elastic energy $E_{tot}/E_0$ = 0.61, 0.60, 0.60, 0.43 for etching depths $h$ = 2.5 µm, 3.4 µm, 5.6 µm and 10.3 µm and tube radius used in Fig. 5. Comparing with the asymptotic total elastic energy of the wrinkled film $E_w/E_0 \cong 0.6$ [11] and of the rolled-up film $E_{R0}/E_0 \cong 0.43$ shows that with increasing etching depth samples go through a transition from wrinkling to tubes. This behavior is distinct from the total elastic energy evaluation from FEM in Fig. 4c.

To understand this difference, one must consider initially that FEM total elastic energy is obtained as an average value for the whole film area. XRD total elastic energy are extracted from measured lattice parameter distribution of an angular section and provide a response which is only locally valid, comprising the tube sector detuned with respect to Bragg's law. In the initial stages of etching, XRD evaluates the strain status of the layer front, which is the only volume of the layers with enough curvature to fulfill Bragg's law locally. This etching front is on a clear wrinkling state as seen by SEM images of Fig. 2a-2c. However, it is clear from FEM calculations on Fig. 4c and SEM measurements that the system already started a rolling process. Integrating through the whole film the elastic energy from FEM leads to $E_{tot}/E_0$ compatible with rolling. The waviness of FEM deformations becomes more pronounced at etching depths of the order of 5 μm, and the normalized elastic energy increases towards higher values, moving to an intermediate wrinkling/rolling state. At such condition XRD analysis is retrieved from a diffraction condition in which the X-ray beam is already impinging the layers at a rolled sector, and a total elastic energy compatible with tubes is retrieved.

The apparent discrepancy described in the previous paragraph can be explained with the fact that FEM calculations cannot properly handle the appearance of cracks, and consequently plastic relaxation, while the total elastic energy calculated from XRD provides a more realistic energetic scenario for $h=$ 10.3 μm, where tubular morphologies are observed by SEM, Fig. 2d. Using the SEM average tube radius of 3 μm for simulating the XRD lattice parameter distribution and in turn the total elastic energy of the sample with $h=$ 10.3 μm, the integrated total elastic energy reduces further, reaching $E_{tot}/E_0 = 0.35$. This value can be considered a lower limit, whereas $E_{tot}/E_0 = 0.43$ (computed with radius 1.6 μm of local sector of tube in Fig. 5) is an upper limit to the XRD total elastic energy of this system at large etching depths. The transition from dominant wrinkling to dominant rolling behaviors, located in between $h=$ 5.6 μm and $h=$ 10.3 μm may depend on the onset of cracks, which has a material-dependent energy of the order of 1 ~ 2eV for GaAs bulk.[42] To our knowledge, failure modes in free-standing nanomembranes are not studied in the recent literature on thin film mechanics.[43] It is expected that nanomembranes have higher strain thresholds for plastic deformation and cracking than bulk materials, indicating that an estimation of the plastic relaxation threshold for our layers may be rather complex. [43,44]

Hence, rolling up tubes near the wrinkling-to-tube transition is therefore an interesting method to produce large strain gradients inside a simple 2-layer system with fixed thickness, also allowing a selection between rolled and wrinkled states. The strain gradient that arises from variations in the local radius can be explored either to probe or use local properties of the rolled layers. Also, the strain gradient might be used to intentionally broaden or shift specific properties in a mesoscopic/macroscopic scale, such as in refs. [2,40]. Our elastic energy evaluations show that XRD is an fairly accurate measurement of local properties, although it cannot be directly used to describe the average behavior of the system. It indicates, nevertheless, that rolling and wrinkling can coexist at a given etching depth, but they'll be found in different sectors of the layers. On the other hand, FEM provides a clear energetic scenario for the beginning of the etching process, since our results integrate the total elastic energy through the released layers full area. However, it does not retrieve the correct energetic state of layers for large etching depths due to the onset of plastic relaxation, clearly observed by SEM and corroborated by XRD. Finally, despite the quantitative differences, by reproducing the layer bending near the etching front and wrinkling at the layers edge, FEM also corroborate the coexistence of wrinkling and rolling status observed by XRD.

In conclusion, we experimentally investigated elastic relaxation of NM on the transition between rolling-up (caused by strain gradient in the NM) and wrinkling (caused by average strain in the NM). Observed NM morphologies indeed show both rolling-up and wrinkling features. Highly focused curvature of wrinkles (strain) in certain points for large etching depth transforms to cracking of the NM in these points. Measured wavelengths of wrinkling scaling linearly with etching depth qualitatively agree with FEM calculations but show large discrepancy with previous analytical scaling predictions. Total strain energy of NMs from FEM calculations confirms more efficient elastic relaxation than pure rolling-up or pure wrinkling. XRD analysis show that the system in the edge of rolling-wrinkling transition can exhibit a large variety of strain status for a fixed layer thickness, being useful to probe how properties change locally with distinct elastic/plastic conditions. The present study sheds light on the interplay of rolling-up and wrinkling for boundary confined films and might serve as platform to induce and control highly localized strains for strain engineering of nanomembranes[27,28] or graphene.[45]

**Supporting Information**. Complementary figures of etching analysis, SEM microscopy and FEM calculations are available. This material is available free of charge via the Internet at http://pubs.acs.org.

AUTHOR INFORMATION

**Corresponding Author**

*peter.cendula@gmail.com

**Present Addresses**

†Institute of Computational Physics, Zurich University of Applied Sciences, Wildbachstr. 21, 8401 Winterthur, Switzerland

§Department of Electrical and Computer Engineering, Faculty of Engineering, Naresuan

University, Phitsanulok 65000, Thailand

**Author Contributions**

The initial layer system was grown by C.D. Lithography, etching experiments and FEM simulations were done by P.C. XRD measurements were conducted and analyzed by A. M. and C. D. The manuscript was written by P. C. and A. M with input from S.K, C.D. and O.G.S. All authors have devised the work and corrected the final version of the manuscript.

**Funding Sources**

@all: Any funds used to support the research of the manuscript should be placed here (per journal style).

ACKNOWLEDGMENT

We acknowledge S. Baunack and E. Coric for SEM images, D. J. Thurmer and D. Grimm for help with photolithography and S. Li for technical assistance.

**Methods**

*Layer growth*

All samples were grown using molecular beam epitaxy, either a a Riber 32p (MPI Festkörperforschung, Stutgart; sample B) or an Omicron system (IFW Dresden, Dresden; sample A). For growth, GaAs (001) substrates were heated and the native oxide thermally removed. After a 200 nm GaAs buffer layer, the actual heterostructure consisting of a 200 nm thick AlAs sacrificial layer and an InGaAs/GaAs bilayer of varying thickness were grown ca. 70°C below the temperature for the change from the c(4x4) to the 2x4 surface reconstruction. Growth was monitored by reflective high energy electron diffraction and the diffraction pattern stayed stricky during the whole growth process indicating a flat surface morphology.

*Sample preparation and etching*

Samples were patterned by optical lithography with long stripe patterns of width 140 μm deep etched with HBr(50% vol.):$K_2Cr_2O_7$ (0.5 mol/L):$CH_3COOH$(100% vol.) (2:1:1) solution (BKC-211). Sacrificial layer thickness was 200 nm for both batches A and B. Etching step of HF was carried out using an HF solution 1:900 (1:450) with etching speed 2.4 μm/min (4.5 μm/min, see Figure S1) for batch A (B) with subsequent immersing in isopropanol and critical point drying to suppress collapse of NMs due to capillary forces. Samples were stored in nitrogen atmosphere before SEM imaging to suppress oxidation.

*Finite element modeling*

We employed commercial finite element software Comsol Multiphysics and its Structural Mechanics Module.[35] Geometrical nonlinearity was turned on to correctly account for large displacements of the bilayer. Epitaxial strain was included as initial strain in the layers and swept from zero to its maximum in typically 100-1000 steps (needed for the convergence). We applied fixed boundary displacement on etching front (supported edge) and symmetric boundary condition on one side (to effectively model larger structure). All other boundaries were free. Small force loads of $10^{-10}$ N were applied in opposite direction at the end points of the bilayer to perturb bilayer from unstable equilibrium state. Tetragonal mesh was used with default density. We refined mesh to check the variation of the obtained solution with mesh resolution. We did not always obtain convergence for refined mesh and the solution was obtained only up to certain initial strain. This hints on the unambiguity of the elastic relaxation of NMs which may eventually lead to self-similar wrinkling patterns observed in similar systems.[7] Our Comsol Model file used in this article is available online.[46]

*X-ray diffraction*

Synchrotron x-ray diffraction measurements were carried out at the XRD2 beamline of the Brazilian Syncrotron Light Laboratory (LNLS). The energy was fixed to 10.1keV and the diffraction was collected by a Pilatus 100K detector mounted with its long axis along the scattering angle, covering an angular range of 4°.